# Elegant Object-oriented Software Design via Interactive, Evolutionary Computation

Christopher L. Simons and Ian C. Parmee
Department of Computer Science and Creative Technologies,
University of the West of England, Bristol, BS16 1QY, United Kingdom
{chris.simons,ian.parmee}@uwe.ac.uk

*Abstract*—Design is fundamental to software development but can be demanding to perform. Thus to assist the software designer, evolutionary computing is being increasingly applied using machine-based, quantitative fitness functions to evolve software designs. However, in nature, elegance and symmetry play a crucial role in the reproductive fitness of various organisms. In addition, subjective evaluation has also been exploited in Interactive Evolutionary Computation (IEC). Therefore to investigate the role of elegance and symmetry in software design, four novel elegance measures are proposed based on the evenness of distribution of design elements. In controlled experiments in a dynamic interactive evolutionary computation environment, designers are presented with visualizations of object-oriented software designs, which they rank according to a subjective assessment of elegance. For three out of the four elegance measures proposed, it is found that a significant correlation exists between elegance values and reward elicited. These three elegance measures assess the evenness of distribution of (a) attributes and methods among classes, (b) external couples between classes, and (c) the ratio of attributes to methods. It is concluded that symmetrical elegance is in some way significant in software design, and that this can be exploited in dynamic, multi-objective interactive evolutionary computation to produce elegant software designs.

## I. INTRODUCTION

DESIGN is fundamental to software development. Indeed, early lifecycle software design is crucial as concepts and information discovered from the design problem domain are used as the basis for many downstream development activities. However, software design is a very human-centered activity, and non-trivial and demanding to perform. Thus in an attempt to assist the software designer in the early stages of the software development lifecycle, evolutionary computation has been increasingly applied using machine-based, quantitative machine-based fitness functions to 'evolve' software designs [1], [2]. Raiha [3] surveys a range of search-based techniques (including evolutionary computing) across a variety of software design activities including both object-oriented and service-oriented architecture design. Bowman *et al.* [4] attempt to solve the class responsibility assignment problem in object-oriented analysis with multi-objective genetic algorithms. Simons *et al.* [5] report the results of evolutionary search supported by interactive software agents in which a population of object-oriented software design individuals is evolved with preference-based designer interaction.

The search techniques reported above rely solely on quantitative computational measures of fitness to direct search and exploration. However, just as evolutionary computing draws inspiration from evolutionary processes in nature, is it also possible to draw from nature to specifically address the 'quality' or 'appearance' of an individual? Certainly, the influence of symmetry of appearance in the reproductive fitness of an organism has been noted by evolutionary biologists. For example, Schilthuizen [6] explains that *"the significance of symmetry was only made clear with the discovery that stress and disease make it harder for an individual to develop a perfectly symmetric body. Small differences on either side of an imaginary mid-plane therefore betray genetic quality, and potential mates use this to gauge each other's desirability. Put simply, symmetry is sexy"*. Drawing from evolutionary biology, it seems likely therefore that symmetry might play a role in subjective designer evaluation of the 'quality' or 'appearance' of a design. Thus it is hypothesized that the symmetrical *elegance* of a software design is a significant qualitative factor in the reproductive fitness of a software design individual and so might be incorporated into the interactive computational evolutionary process. Moreover, previously, Interactive Evolutionary Computation (IEC) has been described as a fusion of evolutionary computation and human evaluation and has been previously applied to a wide variety of areas e.g. arts and animation, music, virtual reality, image processing, data mining, cybernetics, robotics and various other areas [7]. IEC has also been applied with some success in a number of design fields e.g. cartoon facial characters [8], beam bridge design [9], manufacturing plant layout [10], ergonomic chair design [11], and urban furniture such as park benches [12]. However, while such IEC approaches to design exploit subjective evaluation, reports of the use of symmetrical elegance as a measure of design fitness are less readily available in the research literature.

Given the broad research interest generated by IEC, it is interesting to note that the role of elegance in early lifecycle software design is one that hitherto has not stimulated a great deal of research interest. In a more general discussion on software design elegance, Gelernter [13] examines the notion of 'machine beauty' and suggests it can be found *"...in a* *happy marriage of simplicity and power – power meaning the ability to accomplish a wide range of tasks, to get a lot done"*. Indeed, Gelernter discusses the *"the aesthetics of computer science"* and points to the recursive 'quicksort' algorithm as an example of a beautifully designed sorting algorithm that is simple in design yet powerful in performance. Buschmann and Henney [14] do however explore such notions with respect to the design of software architecture. Indeed, Buschmann and Henney begin by posing the question: *"what are the five top properties that make a software design both effective and elegant?"* and go on to suggest that software design economy, visibility, spacing, symmetry and emergence provide *"a perspective on software architecture, a value system that can broadly guide architects' design decisions"*. The authors prudently suggest that it is necessary to balance such considerations however, since the dogged pursuit of one consideration may negatively impact another. For example, too much economy in a design might reduce the overall size, but visibility is lost.

Could symmetrical elegance be significant in software design? Certainly in the 1980s, authors such as Parnas [15] bemoaned a lack of elegance in the software of the era. Parnas pondered why software with a consistent style and simple, organised components was so hard to find, or in other words, *"why software jewels are so rare"*. However, design elegance emerged as a crucial factor in the rise of the software design patterns community. Gabriel [16] discusses patterns of software design and cites Harbison [17]: *"There is a pleasure in creating well-written, understanding software. There is a satisfaction in finding a program structure that tames the complexity of an application. We enjoy seeing our algorithms expressed clearly and expressively. We also profit from our clearly written programs, for they are much more likely to be correct and maintainable than obscure ones"*. The notion of software design elegance helping to tame complexity and enhancing maintainability is pursued further by Gabriel who notes that symmetry among design components is highly useful in defining the dependencies and hence granularity of large scale software designs. Such ideas are directly incorporated by Gamma *et al.* [18] in their seminal design patterns catalogue wherein *"patterns solve specific design problems and make object-oriented designs more flexible, elegant and ultimately reusable"*. Zhao [19] also explores the relationship of symmetry in software design patterns, and suggests that on occasion, it is necessary to break symmetry when applying a design pattern within a specific design problem context.

These ideas are also explored by Wirfs-Brock [20] who ponders the beauty of software design and code. Wirfs-Brock claims that brevity can contribute to code beauty through clarity of purpose and expressive use of the programming language, but only within an elegant design context. Wirfs-Brock also revisits the work of Gabriel, but argues against beauty as an overarching goal in itself. Rather, she suggests that elegance is significant in software design with respect to the development and maintenance of software designs in the face of inevitable change, since elegant designs *"preserve and make evident the designer's intent"*. Wirfs-Brock goes on to invoke Gabriel's notion of software 'habitability', in which software engineers coming to a software design later in its life *"...understand its construction and intentions and change it comfortably and confidently"*. This notion is consistent with the simple conclusion of Tractinsky *et al.* [21] that *"what is beautiful is useable"*.

Although it is generally agreed among the above authors that good software designs exhibit qualities of symmetry and elegance, investigations into the role of symmetry in software design are not readily evident in the research literature. The contribution of this paper, therefore, is to address this shortfall. Indeed, we hypothesize that symmetry and elegance are crucial to software design. Specifically, we further hypothesize that design qualities relating to symmetry and elegance can be exploited within interactive evolutionary computing (IEC), wherein computational fitness and designer evaluation can be combined in a dynamic multi-objective search to lead to elegant, object-oriented software designs.

To achieve this, the nature of the interaction between the human designer and the computational evolution is crucial, and is described in this paper as follows. Section II explains the proposed approach of the investigation, while section III details the experimental methodology followed. Section IV reveals and analyses the results obtained. Threats to validity are discussed in section V, while conclusions are drawn in section VI.

II. PROPOSED APPROACH

This section first provides a brief overview of evolutionary computation used previously in software design. Then, building on this, the proposed approach comprises the following components:
- four novel quantitative elegance measures based on symmetries observed in software designs,
- elegance and reward, and
- dynamic multi-objective evolutionary search

*A. Evolutionary Computation in Software Design*

In evolutionary software design, a population of individual software designs is evolved [5]. The representation of the design solution individual is object-oriented and comprises classes, methods and attributes, consistent with the Unified Modeling Language (UML) [22]. Classes are represented as groupings (or placeholders) of methods and attributes, although, of course, there are many ways in which methods and attributes may be grouped into classes.

The design problem is described by use cases [22], which capture scenarios of interaction between user and the software system-to-be. Within use cases, the steps of the scenarios, and in particular the actions (verbs) and data (nouns) contained in each step, are recorded. If an action and datum are co-located in the same step of the narrative text, the action is said to 'use' the datum. The sets of actions, data and 'uses' thus define the design problem. Design solution attributes are derived directly from members of the set of data specified in the design problem, while methods are derived from members of the set of actions.

The objective fitness function relates to the structural integrity of the software design and measures external coupling between classes in the design solution individual. Referring to the 'uses' of the design problem statement, where a method 'uses' an attribute inside the same class, this is considered to be cohesive. However, where a method 'uses' an attribute outside of its class, an external 'couple' is deemed to exist. It is generally held in software engineering that software designs should strive for high class cohesion and low design coupling. Evolutionary search is thus conducted as a single-objective genetic algorithm using external coupling as a minimization fitness function.

### B. Quantitative Elegance Measures

It is proposed that elegance in software designs is related to the symmetry and evenness of distribution among design elements (albeit in some way that is designer and design context dependent). Therefore, evenness of distribution is quantified though measurements of the distribution of attributes and methods among classes. Four novel quantitative measures of software design elegance are proposed as follows:

1. *Numbers Among Classes (NAC) Elegance* is the standard deviation of the numbers of attributes and methods among the classes of a design and is calculated as follows. Firstly, the average number of attributes per class in a design is calculated together with the standard deviation. Secondly, the average number of methods per class in a design is calculated together with the standard deviation. NAC elegance is calculated as the average of the two standard deviations. The notion here is that the lower the value for NAC elegance, the more symmetrical the appearance of attributes and methods among the classes in the design as a whole.
2. *External Couples (EC) Elegance* is the standard deviation of external couples among the classes of a design and is calculated as follows. For each class in the design, the number of external couples is recorded. The average number of external couples per class is calculated, together with the standard deviation. EC elegance is this standard deviation. The notion here is that the lower the value the EC elegance, the more even the distribution of external couples among design classes.
3. *Internal Uses (IU) Elegance* is the standard deviation of 'internal' uses within the classes of a design and is calculated as follows. For each class in the design, the number of internal uses is recorded. (An internal use occurs when a method in a class 'uses' an attribute in the same class). The average number of internal uses per class is calculated, together with the standard deviation. IU elegance is this standard deviation. The notion here is that the lower the value for IU elegance, the more even the distribution of internal uses among design classes.
4. *Attributes To Methods Ratio (ATMR) Elegance* is standard deviation of the ratio of attributes to methods within the classes of a design and is calculated as follows. For each class in the design, the ratio of attributes to methods is calculated. The average for all ratios is calculated together with the standard deviation. ATMR elegance is this standard deviation. The notion here is that the lower the value of ATMR elegance, the more symmetrical the appearance of attributes and methods in individual classes of the design.

It is evident that all the above elegance measures are minimization functions. It is also noteworthy that while outward indications of NAC, EC and ATMR elegance measures are visible in software design visualizations, IU elegance is not.

### C. Elegance and Reward

It is proposed that the designer interact with the computational evolution as follows. At each generation, external coupling fitness and the four elegance measures are computed. Then, one of the four quantitative elegance measures is selected at random and this measure is used to choose the single most elegant software design solution from the population for visualization. It is proposed that this mechanism provides a means to investigate what elegance measures, if any, are favored by the designer. The designer is then invited to provide a 1 to 5 'elegance' star ranking for the design visualization.

Then, regarding the designer elegance ranking as feedback or 'reward', the mean reward for each of the four elegance measures is calculated as the design episode progresses. Evolutionary search then draws upon this mean reward to dynamically update proportionate selection weights thus producing an interactive, dynamic and multi-objective search. In this way, designer elegance intentions are learned, and the dynamic multi-objective search is steered to reflect those designer elegance intentions.

### D. Dynamic Multi-objective Search

The dynamic multi-objective evolutionary algorithm used in this paper has been inspired by Schaffer's vector evaluated genetic algorithm (VEGA) [23]. In VEGA, the population is divided into equally sized subpopulations for each objective function to be optimized. Each subpopulation is then assigned a fitness value based on a different objective function. After each solution is assigned a fitness value, the selection operator, restricted among solutions of each subpopulation, is applied until the mating pool for the subpopulation is filled. In this manner, restricting the selection operator only within a subpopulation emphasizes good solutions corresponding to that particular objective function. Indeed, since no two solutions are compared for different objective functions, disparity in the ranges of different objective functions does not create difficulty either. However, a number of enhancements are proposed to the VEGA approach to reflect the interactive and dynamic context of evolutionary search. Indeed, it is crucial that the evolutionary search is computationally efficient (so as to not detrimentally impact designer interaction and bring on user fatigue) and yet retain population diversity. Because of this, separate subpopulations are not maintained. Rather, a single population of software design solutions and a proportionate selection operator is proposed. Building on results of previous studies ([5]), the selection operator used is binary tournament selection. However, the tournament selection operator uses the appropriate fitness function (i.e. external coupling or one of the four elegance metrics) for tournament comparison in proportion to the dynamic elegance 'reward' received from the designer as the evolutionary search progresses.

Let the interactive designer elegance ranking be regarded as 'reward', $r$, for the four elegance metrics, $r_{e_1}, r_{e_2}, r_{e_3}$ and $r_{e_4}$ respectively. A mean reward for each elegance measure is computed based on the sequence of designer interactive evaluations during the design episode as follows

$$\bar{r}_{e_1} = \frac{1}{n} \sum_{k=1}^{n} r_{e_1,k} \quad (1)$$

where $n$ is the number of interactive rewards provided by the designer. The mean reward values for the other elegance values $\bar{r}_{e_2}, \bar{r}_{e_3}$ and $\bar{r}_{e_4}$ are similarly defined. Let the selection weights of the five quantitative measures be $w_c$, the weight for external design coupling, and $w_{e_1}, w_{e_2}, w_{e_3}, w_{e_4}$ the weights for the four elegance measures respectively. The scale of the selection weight for each elegance measure is equal i.e.

$$0.0 \leq w_{e_1}, w_{e_2}, w_{e_3}, w_{e_4} \leq 0.2$$

The selection weight for external coupling is calculated as follows:

$$w_c = 1.0 - (w_{e_1} + w_{e_2} + w_{e_3} + w_{e_4}) \quad (2)$$

Thus when the selection weights for each elegance metrics are at maximum values (i.e. 0.2), the selection weights for all fitness functions are equal.

At the beginning of evolutionary search prior to designer interaction, the selection weight for each elegance measure is zero and so external coupling is used as the sole selection metric over the entire search population. However, as search progresses, the designer provides elegance evaluations of designs for an elegance measure chosen at random, from which the mean reward for the chosen elegance measure is updated. The selection weights for each elegance measure are then updated as follows:

$$w_{e_i} = \bar{r}_{e_i} \cdot c \quad (3)$$

A value of 0.04 is used for the constant, $c$, as this effectively maps the scale of the 1 to 5 reward star ranking to the upper value of the elegance weightings of 0.2. At each designer interaction, the weighting for external coupling is also updated as in equation (2). In this manner, the dynamic multi-objective search emphasizes quantitative external coupling at the start of search, but as designer interactions increasingly contribute to search, the selection weightings of elegance measures increase. Indeed, should the designer reward one elegance measure above all others, search is increasingly steered to design solutions reflecting this elegance measure. The interaction selection weighting update mechanism is summarized as follows.

1) Randomly select an elegance measure, $e_i$, from elegance 1..4
2) Select most elegant design from population using $e_i$
3) Present visualization of elegant software design
4) Obtain designer ranking in range 1 to 5 (star rating)
5) Update mean reward $\bar{r}_{e_i}$ for $e_i$
6) Update selection weighting $w_{e_i}$ based on mean reward $\bar{r}_{e_i}$
7) Update selection weighting $w_c$

### III. METHODOLOGY

Three software design problem domains are used as vehicles for investigation. The first software design problem domain is a generalized abstraction of a Cinema Booking System (CBS), which addresses, for example, making an advance booking for a showing of a film at a cinema, and payment for tickets on attending the cinema auditorium. A specification of the use cases of Cinema Booking System design problem is available at [24]. The second software design problem domain is an extension to a student administration system performed by the in-house information systems department at the University of the West of England, UK. The university sought to record and manage outcomes relating to the Graduate Development Program (GDP) of students during their studies. The extension was implemented and deployed in 2008. A specification of the use cases used in the development is available from [25]. The third software design problem domain is based on an industrial case study – Select Cruises (SC) - relating to a cruise company selling nautical adventure holidays on tall ships where passengers are members of the crew. The resulting computerized system handles quotation requests, cruise reservations, payment and confirmation via paper letter mailing. A specification of the use cases of Select Cruises design problem is available at [26]. With respect to the design problem scale, table 1 shows the number of attributes, methods and uses for each design problem.

Manual software designs have been performed by the appropriate software engineers for the three problem domains as an integral part of the case study activities and are available at [27].

TABLE 1: Scale of Software Design Problems.

| Design Problem | Number of Attributes | Number of Methods | Number of Uses |
|---|---|---|---|
| CBS | 16 | 15 | 39 |
| GDP | 43 | 12 | 121 |
| SC | 52 | 30 | 126 |

5TABLE 2. Trial Participant Information.

| Participant | Gender | Current Profession | Years in Industry | Years in Academia | Total Years |
|---|---|---|---|---|---|
| 1 | male | Software Engineer / Lecturer | 12 | 10 | 22 |
| 2 | male | Undergraduate Student | 1 | 2 | 3 |
| 3 | male | Lecturer | 12 | 23 | 35 |
| 4 | male | Software Engineer / Researcher | 29 | 4 | 33 |
| 5 | female | Lecturer | 0 | 10 | 10 |
| 6 | female | Lecturer | 0 | 20 | 20 |
| 7 | female | Lecturer | 5 | 20 | 25 |
| | | Total | 59 | 89 | 148 |

Seven software development professionals with experience of early lifecycle software design participated in trials using the approach proposed in the previous section. Relevant information concerning the seven software professional is given in table 2. The total experience of software development of the participants amounts to 148 years in both academia and industrial practice. Participant 1 is the first author of this paper.

All participants engage with the same interactive evolutionary computational environment. Prior to interacting with the evolutionary search, each participant receives the same briefing of each example software design problem, and the colorful visual designer interface is described. The participant is not informed of which elegance measure has been chosen at random to select a software design for visual inspection; the participant simply performs their qualitative ranking of perceived design 'elegance'. Once underway, the interactive design episode is allowed to proceed until the participant decides to halt.

At each designer interaction, the following are recorded:
- external coupling,
- NAC, EC, IU and ATMR elegance measures,
- designer qualitative ranking of design elegance (values range from one 'star' to five 'stars'),
- updated mean reward for each elegance measure, and
- updated selection weights for external coupling and each elegance measure.

Lastly, each participant is invited to provide any comments on their overall human experience of the trial. Such comments might include any satisfying aspects, any aspects that generated user fatigue, and any suggestions for enhancement of the overall human experience.

## IV. RESULTS

Screen shots of example early lifecycle software designs illustrating what might be considered examples of elegant and inelegant designs for the Cinema Booking System, Graduate Development Program and Select Cruises can be found at [28]. In this section, results of investigations into external coupling fitness are presented first, followed by results relating to then designer evaluation and reward. Next, results of dynamic multi-objective search are revealed while lastly, participant comment on the human experience of the interactive framework is presented. A total of 149 interactive evolutionary design episodes have been conducted with the participants, involving 1,942 designer interactions in total. Experimental results data can be found at [29].

### A. External Coupling

Values of average population external coupling achieved at designer halting after evolutionary search for each of the software design problems are shown in table 3, together with external coupling values of the manual software designs for comparison. Average population external coupling are broadly similar to the values obtained previously with the manual designs. Nevertheless, interpretation of these computational quantitative findings must be considered in the light of qualitative designer evaluation and its impact on evolutionary search. Therefore results of experiments into elegance and designer reward, and dynamic multi-objective search are described in the following section.

TABLE 3. Average Population External Coupling values achieved after Evolutionary Search.

| Design Problem | Manual Design | External Coupling | Standard Deviation |
|---|---|---|---|
| CBS | 0.154 | 0.202 | 0.019 |
| GDP | 0.297 | 0.305 | 0.041 |
| SC | 0.452 | 0.449 | 0.082 |

## B. Elegance and Reward

Values of average population elegance achieved at designer halting after evolutionary search for each of the software design problems are shown in table 4, together with elegance values of the manual software designs for comparison. Superior elegance values are shown in bold. The manual software designs can be regarded as baseline designs with respect to cohesion of classes and coupling among classes in the designs. However, it is interesting to note that average NAC and IU elegance is superior for the interactively evolved designs in all three design problems. ATMR elegance is also superior for the two larger design problems i.e. GDP and SC. Only EC coupling is superior in all three manual design problems. This suggests that symmetrical elegance as measured by NAC, IU and ATMR could be an important factor in the interactive evolution of the three software designs, whereas EC elegance could be less important. Thus to investigate if any possible relationships exist between elegance measures, elegance values have been recorded during interactive designer episodes.

Figure 1 reveals an example of a design episode for the Cinema Booking System in which values for each elegance measure is shown. No attempt has been made to normalize the measures - raw data are plotted as recorded. The example provided is considered sufficiently representative of the bulk of design episodes drawn from the three example design problems; for the sake of brevity, further examples have not been included. It is apparent from figure 1 that considerable variability exists in the fitness of the quantitative elegance measures as the design episode progresses. However, in multi-objective dynamic search, elegance fitness measures can conflict. Thus as can be seen at later generations in figure 1, although EC elegance increases, IU elegance deteriorates.

To investigate the variability in the distribution of quantitative elegance measures, further analysis has been conducted. For each of the 149 design episodes, best in population values for the four elegance measures have been recorded at designer halting of interactive evolutionary search.

TABLE 4. Average Population Elegance values achieved after Evolutionary Search.

| Problem | Elegance | Manual Design | Average Elegance | Standard Deviation |
|---|---|---|---|---|
| CBS | NAC | 0.821 | **0.641** | 0.508 |
|  | EC | **0.894** | 1.003 | 0.485 |
|  | IU | 3.429 | **2.160** | 1.011 |
|  | ATMR | **0.199** | 0.261 | 0.194 |
| GDP | NAC | 2.592 | **0.967** | 0.524 |
|  | EC | **2.712** | 3.823 | 1.964 |
|  | IU | 15.263 | **6.427** | 2.783 |
|  | ATMR | 2.617 | **1.439** | 0.806 |
| SC | NAC | 1.520 | **0.642** | 0.326 |
|  | EC | **2.471** | 3.316 | 0.728 |
|  | IU | 4.608 | **2.225** | 0.948 |
|  | ATMR | 1.848 | **0.948** | 0.350 |

The mean values for each elegance measure are as follows:
- NAC Elegance: 1.63
- EC Elegance: 3.19
- IU Elegance: 3.61
- ATMR Elegance 4: 1.56

At first glance, it seems that mean values for NAC and ATMR elegance are superior to those for EC and IU elegance. Thus to ascertain if the differences between the distributions of the elegance values are statistically significant, the Friedman test is applied and reveals that $p < 0.01$, indicating that the rankings differ with statistical significance across the elegance measures. To further investigate if the differences between distributions of pairs of individual elegance measures are significant, a Wilcoxon matched-pairs, signed-rank test is conducted. This test reveals that the differences in the distribution of rankings between all pairs of elegance measures are significant beyond the .01 level ($p < 0.01$), except for the pairing NAC elegance and ATMR elegance ($p = 0.129$). This suggests that while generally there is significant variability between the four quantitative elegance measures, the distributions of NAC elegance and ATMR elegance may be similar.

Reward rankings obtained at designer halting of each interactive evolutionary episode have also been recorded and analyzed. The mean rankings for each reward are:
- NAC Reward: 2.53
- EC Reward: 2.33
- IU Reward: 2.47
- ATMR Reward: 2.67

To ascertain if the differences between the distributions of the reward rankings are statistically significant, the Friedman test is again applied to reveal that $p = 0.108$, indicating that the rankings do not differ with statistical significance across the reward measures. This suggests that unlike the results for elegance values, the differences between reward rankings are not significant. However, it is instructive to recall that other authors (e.g. [7], [9]) have reported the possibility of loss of linearity of participant focus over the trajectory of time during an interactive evolutionary computing episode. Thus it is also possible that designer qualitative ranking itself might at times be inconsistent, which is perhaps not surprising for such an abstract concept such as design elegance. Furthermore, it is likely that qualitative evaluation depends on the overall human experience of interaction, which includes any pre-existing personal experience and perspective that the designer brings to the situation. To analyze the reward results further, application of the Wilcoxon signed-rank test between individual pairs of reward rankings may also be useful insofar as it is conditional upon where change has occurred (i.e. ranking scores that tie are removed from the calculation of significance). The results of this test suggests there are no statistically significant differences between individual pairs of reward rankings, except for the rankings of reward obtained for EC reward and ATMR reward 4, albeit at the $p < 0.05$ level. This suggests that overall, there is little significant variability between distributions of the four reward rankings.

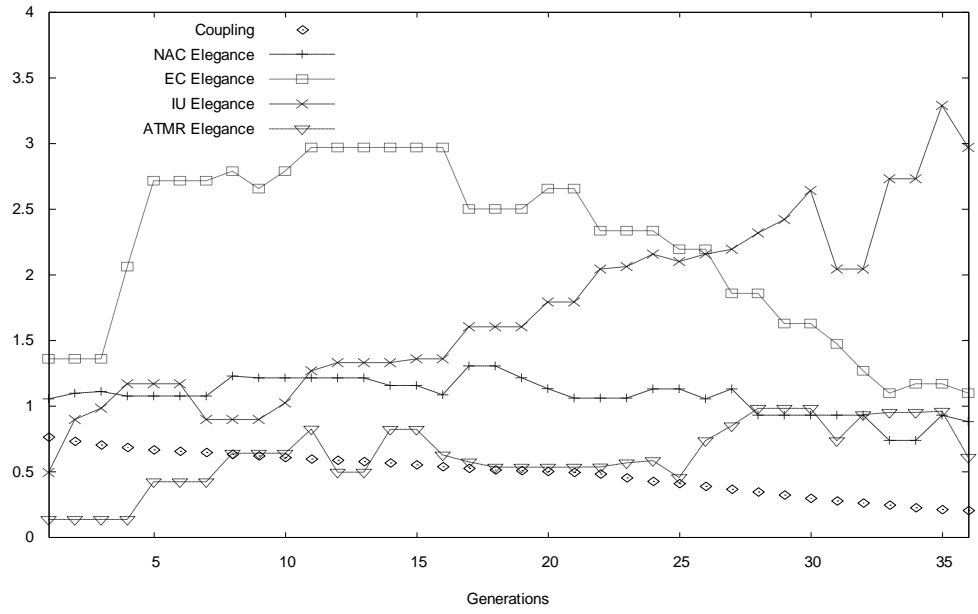

Fig.1. Example of Quantitative Elegance Measures during a
Single Design Episode for Cinema Booking System

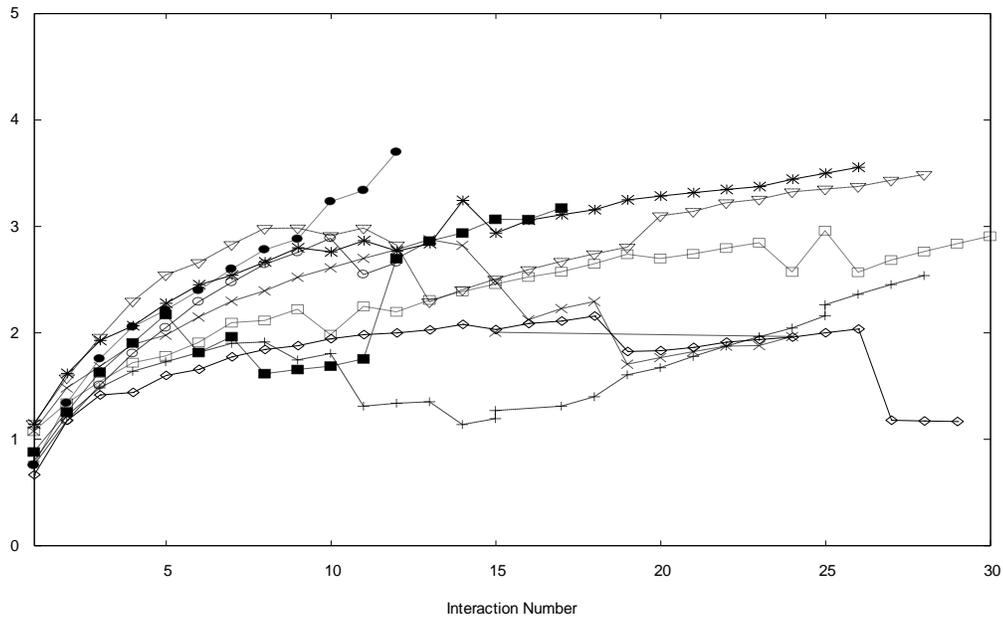

Fig.2. Average Reward at each Interaction for Various
Design Episodes with Cinema Booking System

TABLE 5. Spearman's Rank Correlation between Elegance and Reward

|  |  | NAC Elegance | EC Elegance | IU Elegance | ATMR Elegance |
|---|---|---|---|---|---|
| NAC Reward | Correlation Coefficient | -0.264 | -0.305 |  | -0.287 |
|  | Sig. (2-tailed) | **.001** | **.000** |  | **.000** |
| EC Reward | Correlation Coefficient | -0.211 | -0.264 |  | -0.234 |
|  | Sig. (2-tailed) | .015 | **.002** |  | **.007** |
| IU Reward | Correlation Coefficient | -0.220 | -.344 |  | -0.307 |
|  | Sig. (2-tailed) | .012 | **.000** |  | **.000** |
| ATMR Reward | Correlation Coefficient | -0.228 |  |  | -0.243 |
|  | Sig. (2-tailed) | .011 |  |  | **.007** |

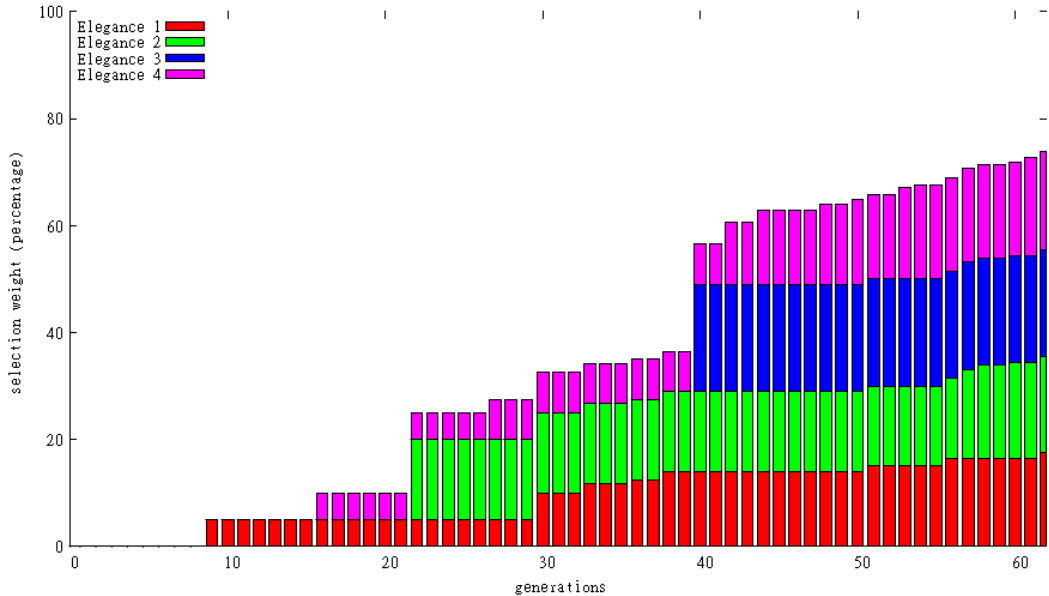

Fig. 3. Stacked Histogram of Typical Elegance Selection Weights for a Single Design Episode with the Cinema Booking System Design Problem

TABLE 6. Evolutionary Generation Selection Weightings at Design Halting for Example Single Design Episodes

|     | Generation | $w_{e_1}$ | $w_{e_2}$ | $w_{e_3}$ | $w_{e_4}$ | $w_c$ |
|-----|------------|-----------|-----------|-----------|-----------|-------|
| CBS | 62         | 0.175     | 0.180     | 0.200     | 0.183     | 0.262 |
| GDP | 69         | 0.108     | 0.150     | 0.200     | 0.200     | 0.342 |
| SC  | 187        | 0.129     | 0.125     | 0.150     | 0.183     | 0.413 |

At this point, it is interesting to speculate: what relationships, if any, exist between elegance and reward? Thus to investigate the strength of any association between elegance and reward, Spearman's rank correlation is applied and the results are shown in table 5. For clarity, only statistically significant correlations are shown; correlations significant at the 0.01 level (2-tailed) are shown in bold. It is immediately evident from table 5 that IU elegance does not correlate with any of the reward rankings. In addition, EC elegance does not correlate with ATMR reward rankings. However, there is a highly statistically significant correlation between ATMR elegance and all four reward rankings. Furthermore, there is also a statistically significant correlation between NAC elegance and all four reward rankings, albeit at the p < 0.05 level for EC, IU and ATMR rewards. Lastly, there is also a highly statistical significance between EC elegance and NAC, EC and IU rewards. Overall, therefore, with the notable exception of IU elegance, the negative correlation coefficient values suggest a degree of association between NAC and ATMR elegance values and reward rankings. In other words, as NAC, ATMR and (to a lesser extent) EC elegance values decrease (i.e. become superior), increasing reward is obtained from the software designer. This appears to be in-keeping with the previous finding that NAC and (to a lesser extent) ATMR elegance values for software designs produced by interactive evolution are superior when compared with the baseline manual software designs. Taken together, this evidence suggests that elegance measures based primarily on class symmetry (i.e. NAC and ATMR elegance) play some significant role in the interactive design of UML class models, whereas the role of other measures (such as EC and IU elegance) is less important. It would appear that the evenness of distribution of external couples is less significant than class-based symmetry, and evenness of internal use distribution is not significant at all.

### C. Dynamic Multi-objective Search

During interactive software design episodes, dynamic selection weightings have been recorded and results of a typical single individual episode for the Cinema Booking

System are shown in figure 3. Values of selection weightings for typical single episodes for each design problem at designer halting are given in table 6. It is interesting to note that figure 3 reveals slowly changing selection weights for each elegance measure from approximately generation 50 to designer halting. This is because the average reward obtained from qualitative designer evaluation has been found to be slowly changing during these generations too. Therefore, taken as a whole, the results presented in figure 3 and table 6 appear to reveal a picture of dynamic multi-objective interactive evolutionary search wherein the selection weightings of elegance measures respond in a timely manner to the reward obtained from the software designer.

### D. Human Experience

Participant designers were invited to provide any comment on their human experience of the interactive evolutionary design episodes. Four of the participants took up the invitation; their comments are provided in full at [30] and an analysis is provided as follows. The participants appear to have regarded the software designs visualized for evaluation as natural in appearance. Indeed, overall, the participants found the interactive software design episodes to be very engaging. Participant 3 reports that *"I found the sessions quite enjoyable, the enjoyment gained from working towards, and seeming to achieve, a useful goal: i.e. a good design"*. It is interesting to note that at times, participant 3 felt that they were attempting to 'encourage' the interactive framework to design discovery by providing reward. Participant 7 reports that *"I found the tool to be surprisingly engaging, so much so that, at times, I think I lost sight of the aim of the task and was more focused on the looking to see how the results changed from one output run to the next"*.

The participants also report the effectiveness of the graphical visualizations of software designs, and highlight the use of color. Participant 5 remarks that *"the use of color was a heavy influence"*. Participant 7 comments that *"... colors had a huge impact on my decision making"*. Participant 6 reports that with respect to the graphical interface of the interactive framework, *"the interface of the tool is very easy to follow. The idea of being able to visualize the degree of cohesion and coupling is very good. I believe the tool helps the users to easily understand the quality of the software design."* Nevertheless, despite the effectiveness of software design visualization, the participants have also provided insights into the relationship between the abstract concept of design elegance and reward. For instance, Participant 2 reports that *"I'm not convinced that I really made any 'elegance' judgments – my judgment was principally guided by the tool"*. Participant 5 also comments on feeling uneasy and daunted by judgments of design elegance: *"I felt that sometimes that my judgment values altered during the course of a run – especially for the more complicated examples. So... [sic] the perceived lack of consistency could undermine the confidence in the value of the decisions"*. Such a lack of consistency in elegance judgment may go some way to explain the variability in reward rankings.

### V. THREATS TO VALIDITY

The outcomes achieved in the study depend highly on the design problems used. The second and third design problem domains (Graduate Development Program and Select Cruises) are taken from authentic industrial developments and so provide examples of design problems of industrial scale and complexity.

The interactive evolutionary design experience is also highly dependent on the design context, and so every attempt has been made to make a consistent design context for all participants. The same briefing has been received by all participants and all trials have been conducted in the same interactive evolutionary computational environment.

The outcomes of the investigations also depend heavily on the number and experience of the participants. The 148 years' experience of professional software development among the seven participants in the experiments includes 89 years of academic experience. It also includes 59 years of industrial software design and development experience for participants 1, 3 and 4 who have architected and developed software across a wide variety of software design domains, within object-oriented and service-based technical architectures worldwide. While a greater number of participants would have lent greater robustness to the study, the years of experience of the trial participants suggests a level of credibility for their elegance evaluations.

In addition, the results of the study depend upon the number of recorded interactions made by the participants with the interactive evolutionary design. To address this, participants engaged in 149 design episodes comprising 1,942 designer interactions in total.

### VI. CONCLUSIONS

For three out of the four novel elegance measures proposed, there appears to be a significant correlation between the quantitative elegance value of a software design and the subsequent reward elicited from the designer. We conclude therefore that it is likely that symmetrical elegance is in some way significant in software design. The three elegance measures that appear to correlate with designer reward rankings are NAC elegance, EC elegance and ATMR elegance, suggesting that evenness of distribution of class design elements in particular is a factor in stimulating subsequent reward from the designer. The elegance measure that does not correlate with designer reward is IU elegance. We speculate that this is because IU elegance has no discernible visual impact on software design appearance, whereas the other elegance measures do.

It is also an interesting finding that designer reward rankings for each elegance measure appear to be similar in distribution, possibly suggesting that designers are in their minds evaluating a more abstract, general quality (or qualities) of software design elegance, rather than attributing design elegance to a single specific measure. Of course, it is possible that the designers are providing reward based on a measure other the four measures proposed in this study. However, the significant correlation between NAC, EC and ATMR elegance and designer reward rankings does suggest these measures play some part in determining reward (although significant correlation shows association but does not imply causation).

Participant comment on their interactions with interactive evolutionary computation is highly positive, with a number of participants stating how much they enjoyed the interactive software design experience. Overall, experimental results and participant comment taken together suggest that the novel quantitative symmetrical elegance measures relating to numbers among classes (NAC), external couples (EC) and attribute to method ratio (ATMR) are in some way significant in software design and so can indeed be exploited in an effective dynamic, multi-objective interactive evolution to produce elegant software designs.